# Scatterer assisted whispering gallery mode microprobe

Fangjie Shu,[1,2] Xuefeng Jiang,[1] Guangming Zhao,[1] and Lan Yang[1,*]

[1]*Department of Electrical and System Engineering, Washington University in St. Louis, St. Louis, Missouri 63130, USA*
[2]*Department of Physics, Shangqiu Normal University, Shangqiu 476000, P. R. China*
*\*Corresponding author: yang@seas.wustl.edu*

Fiber based whispering-gallery-mode (WGM) microprobe, combining both the high optical field enhancement of the WGMs and the practicability of the fiber probe, is highly demanded in sensing and imaging. Here in this paper, we experimentally report the efficient far-field coupling of WGMs by scattering the focused laser beam through a nanotip. With the help of Purcell effect as well as the two-step focusing technique, a WGM excitation efficiency as high as 16.8% has been achieved. Both the input and output of the probe light propagate along the same fiber, which makes the whole coupling system a fiber based WGM microprobe for sensing/imaging applications. © 2017 SIOM

## 1. INTRODUCTION

Whispering gallery mode (WGM) microresonators have attracted increasing attentions in the last two decades in the fields of microlaser [1-5], optomechanics [6-8], bio/chemical/thermal sensing [9-17], nonlinear optics [18-23], *etc.*, due to their intense ligh confinements originating from ultrahigh quality ($Q$) factors and small mode volumes. It is critical to effectively couple light into and out of the microresonator for the aforementioned experiments and applications. Traditionally, a near-field coupler, *e.g.*, an integrated waveguide or a tapered fiber, within the evanescent field of the resonant mode, is used to couple light into and out of the WGM microresonator [24, 25]. However, to get the record coupling efficiency the evanescent field coupling method essentially requires not only the strict phase-matching condition [26], but also a high-resolution lithography to define the wavelength-scale waveguide-resonator gap or a high-precision alignment of the fiber taper. Besides, cantilever-type evanescent field coupler, such as a fiber taper, suffers from the mechanical vibration, which gives rise to the instability of the coupling loss as well as the storage energy in the resonator. Another approach is to tailor the cavity geometry to achieve a directional output [27-30], and thus the cavity modes can be time-reversely excited by a free-space laser beam [4, 31-36]. Although some excellent designs have produced a unidirectional emission with a relatively small divergence angle and achieved the free-space coupling successfully, the coupling efficiency is much lower than the near-field coupler, especially for the ultrahigh $Q$ factor modes [32]. Besides, each cavity material (refractive index) requires a particular cavity shape design, making it inconvenient to operate.

Another alternative coupling method is scattering the free-space laser beam into resonant modes by an extra defect/scatterer placed inside the mode field [37-39]. This coupling method is free from the phase-matching condition and suitable for resonators with different materials. However, a practical coupling scheme with a high coupling efficiency and a compact coupler is still unexplored so far. Here in this paper, for the first time we experimentally realized the efficient far-field light coupling between a conventional single-mode optical fiber and high-$Q$ WGMs via a scatterer placed on the surface of a microsphere resonator. Both input and output lights are transmitted along the same optical fiber. Specifically, a nanotip mounted on a 3-axis nano-stage is utilized as a light scatterer, so we could adjust its position on the resonator in the experiments. A graded-index (GRIN) lens coupled with a single mode fiber is used to focus the laser beam in 1,550 nm wavelength band onto the nanotip, which then scatters the focused laser into WGMs with a coupling efficiency as high as 16.8%. Particularly, the reflection signal is also collected by the same lens and fiber, which makes the whole coupling system a fiber based WGM microprobe, which will become a compact platform for sensing/imaging applications.

## 2. THEORIES AND EXPERIMENTS

### A. THE COUPLING PROCESS AND THE SPECTRUM

The experimental setup, as shown in Fig. 1(a), consists of a tunable laser, a circulator, a fiber coupled GRIN lens, a microsphere resonator with a diameter of 35 μm, a fiber nanotip with a diameter of around 0.2 μm at the top, and a photodetector. The probe laser in 1,550 nm wavelength band is first coupled into the fiber and then focused onto the surface of the microsphere by the GRIN lens with a working distance of 200 μm. Different from the traditional focusing method, here we propose a special two-step focusing technique. Specifically, the focused laser beam with a size of around 6 μm is further focused by the microsphere itself onto the nanotip located at the back surface of the microsphere. Most of the light scattered by the nanotip is coupled into the resonant mode with the help of Purcell effect when the incident light is scanned across a WGM. Remarkably, the incident light coupled equally to the clockwise (CW) and counter-clockwise (CCW) modes benefitting from the Rayleigh scattering of the nanotip. Meanwhile, the cavity mode field can also be coupled out of the cavity by the same nanotip in the reversed optical path and then collected by the same GRIN lens and fiber. Finally, the collected reflection signal is routed by the circulator and detected by the photodetector, which is connected with an oscilloscope to monitor the back-scattering spectrum of the resonant system. In the experiments, three 3-axis nano-stages are used to control the relative positions of the microsphere, the nanotip, and the incident laser beam to adjust the coupling condition.

A typical reflection spectrum is shown as the black curve in Fig. 1(b). As a comparison, the red curve in Fig. 1(b) shows the reflection spectrum when the nanotip is removed from the surface of the microsphere, and the resonant peaks disappear subsequently, from which we can conclude that the resonant peaks are caused by the Rayleigh scattering of the nanotip. Note that the doublet peak exhibits a clear mode splitting, which is caused by the mode coupling of the CW and CCW modes induced by the scattering of the nanotip [10]. It further proves the role of the nanotip in the Rayleigh scattering. In other words, the nanotip in the coupling system not only couples resonant light into and out of WGMs of the microsphere but also provides mode coupling between the CW and CCW modes. Moreover, this coupling mechanism is free from the phase-matching condition, and applies to all the modes in the microsphere. As shown in Fig. 1(c), there are five spikes in the reflection spectrum (red curve), showing the same frequencies with resonant dips in the transmission spectrum (black curve), which is obtained by the traditional fiber taper coupling method. It is worth noting that some modes are not excited by the nanotip, which can be attributed to the mismatching between the mode distributions and the location of the nanotip.

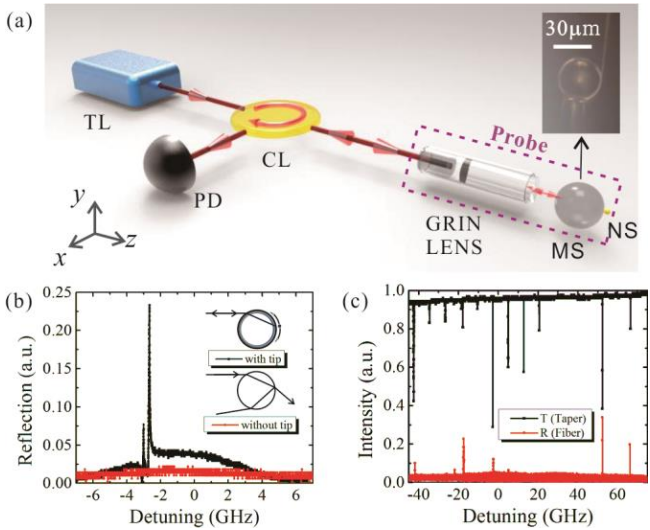

Fig. 1. (a) Illustration of the experimental setup. TL: tunable laser, PD: photo-detector, CL: circulator, GRIN LENS: graded-index lens, MS: microsphere, NS: nano-scatterer. Inset: an optical image of a microsphere and a fiber nanotip in the experiment. (b) Reflection spectra with (black) and without (red) a nanotip. (c) Optical spectra of the scatterer-assisted coupling (red) and the fiber taper coupling (black).

An overall coupling efficiency of 2.8% is derived from the measured reflection signal peak power divided by the input probe power. Note that the overall coupling efficiency includes both the input and the output coupling processes, which are two reversible processes and thus should have the same coupling efficiency. Therefore, the coupling efficiency of the proposed coupling method is the square root of the overall coupling efficiency, *i.e.*, 16.8%. The high scattering-based coupling efficiency is attributed to both Purcell effect and the proposed two-step focusing technique.

## B. COUPLING EFFICIENCIES AS A FUNCTION OF THE POSITIONS OF THE LENS

The free-space coupling efficiency is typically determined by the beam matching between the resonant emission pattern and the incident probe laser beam [31]. To obtain a higher coupling efficiency, one need to engineer the free-space Gaussian beam to have a better overlap with the nanotip induced emission pattern, including both the divergence angle and the coupling position. To measure the divergence angle of the nanotip induced emission pattern, we excite the resonant mode by a tapered fiber and measure the 3-dimension emission pattern by moving the 3-axis nano-stage, where the GRIN lens is mounted, as shown in Fig. 2. A divergence angle of 4.18° is obtained by fitting the focused profile of the emission pattern. In the experiment, a GRIN lens with a working distance of around 200 μm is utilized to focus the probe laser beam down to around 6 μm. The divergence angle of the focused beam is about 3.27°, which is similar with that of the emission divergence.

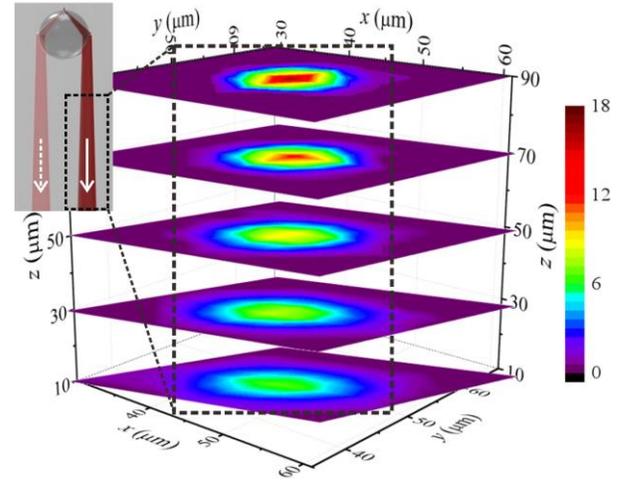

Fig. 2. Measured 3-dimension nanotip induced emission pattern. Inset: a sketch shows the position of the emission pattern.

To study the dependence of the coupling position, we mount the GRIN lens, the microsphere, and the nanotip on three 3-axis nano-stages, respectively. By moving the GRIN lens solely, we investigate how the coupling efficiencies of a $Q = 5.6 \times 10^6$ mode vary with the position of the incident light (Fig. 3), with the coordinate system defined in Fig. 1(a). The maximum reflection intensities, which indicate the coupling efficiencies, as a function of the relative $x$ and $y$ between the probe laser beam and the microsphere are shown in Figs. 3(a) and 3(b), respectively. Reflection peaks exist when adjusting both $x$ and $y$ positions of the GRIN lens. The width of the peaks at the half height are 3.2 μm and 7.1 μm for the relative $x$ and $y$, respectively. In other words, the coupling system has a position tolerance of about 2 to 4 times of the wavelength in the lateral plane. Note that the $y$-related width is more than two times of that in the $x$ direction, which is because the nanotip is a scatterer with a larger dimension in the $y$ direction than in the $x$ direction. Different from $x$ and $y$ positions, there is a periodic oscillation of the reflection signal with varied $z$ positon, as shown in Fig. 3(c). The period of the oscillation is about 0.77 μm, which is the half of the probe light wavelength. This oscillation stems from the interference between the reflected light from the end-face of the GRIN lens and the resonant emission of the microsphere. The average reflection signal decreases less than 20% in a range of 10 μm, showing a very large position tolerance in the $z$ direction. It is worth noting that the $x$-related curve is asymmetric in Fig. 3(a), which is attributed to the non-resonant reflection of the microsphere. Specifically, at the position A, where the input light aims at the center of the microsphere, the non-resonant reflection of the probe light by the microsphere can be collected by the coupling lens, then forms a high reflection background, as shown in the frequency-scanning curve in Fig. 3(d). While at the position C, where the probe beam is far away from the microsphere, neither resonant nor non-resonant light can be reflected.

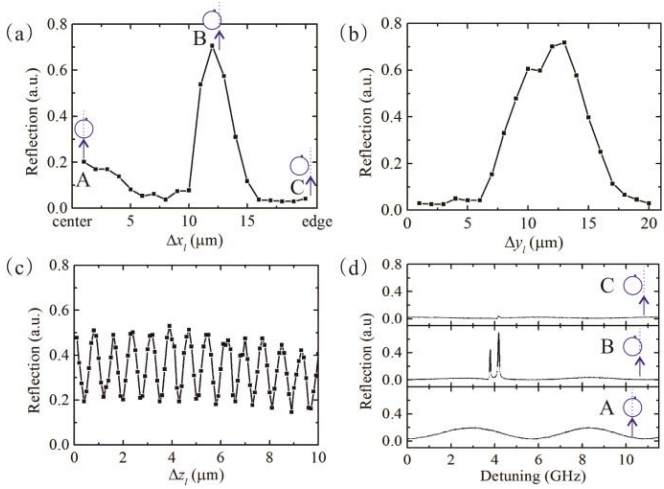

Fig. 3. Peaks of reflection spectrum vary with the relative position $\Delta x_l$ (a), $\Delta y_l$ (b), and $\Delta z_l$ (c) of the GRIN lens (incident beam) when keeping the nanotip-sphere coupling system invariant. (d) Reflection spectra at three positions of $x$.

## C. COUPLING EFFICIENCIES AS A FUNCTION OF THE POSITIONS OF THE NANOTIP AND THE MICROSPHERE

In this section, we investigate the coupling efficiency varying with the relative position of the nanotip on the equator of the microsphere. The experiment is performed by moving both the nanotip and the microsphere in the $x$ axis with the incident laser beam focused. More specifically, the microsphere moves with a step of 2 μm. For every position of the microsphere, we move the nanotip along the equator of the microsphere in the $x$-$z$ plane continuously and record both the top-view imaging by a CCD camera and the reflection spectrum with a frame rate of ~ 3 Hz.

The position of the nanotip on the microsphere is indicated by a polar coordinate, $\theta$, in the $x$-$z$ plane, as defined in the left inset of Fig. 4(a). Specifically, when the nanotip is near or on the equator of the microsphere, a light spot caused by the resonant scattering of a illumination light can be imaged in the top view, which can help to mark the position of the nanotip, $\theta$. Meanwhile, the reflection signal, which corresponds to the coupling efficiency of every nanotip position, is acquired by a photodetector. Figure 4(a) shows the reflection as a function of the nanotip position with the microsphere fixed at $\Delta x_s = -11$ μm, where a main peak accompanied with several side lobes is presented. The width of the main peak is about 3 degree which corresponding to 0.9 μm. A 2-dimension simulation with a Gaussian beam focused by a microsphere is performed to explain the fringes. The simulated field distribution around the rim of the microsphere is shown in the right inset of Fig. 4(a), where several fringes can be clearly seen on the surface of the microsphere. Note that the size of the nanotip (~ 0.2 μm) is much smaller than the period of the fringes (~ 1 μm). Thus the fringes can be resolved by the nanotip moving along the ring of the spherical resonator. In addition, the fluctuation of the coupling efficiency maps the distribution of the intensity of the focused field, which further verifies that the resonant coupling is caused by the nanotip.

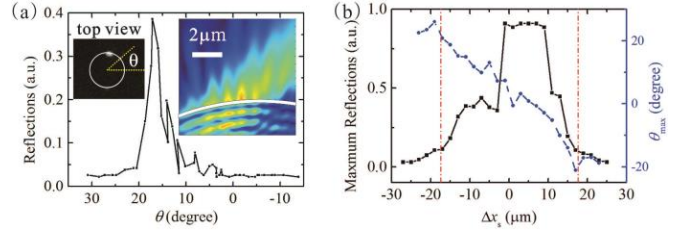

Fig. 4. (a) Peak value of the reflection varies with the position of the nanotip on the sphere with the microsphere fixed at $\Delta x_s = -11$ μm. Left inset: a top view of the sphere-tip coupling system. Right inset: the simulated distribution of a Gaussian beam focused by the microsphere. (b) Maximum peak value in the reflection spectrum (black solid curve) and the corresponding position of the nanotip (blue dashed curve) vary with the coordinate of the sphere, $\Delta x_s$, when the incident beam is fixed. The vertical red dot dashed lines indicate the boundaries of the microsphere at $\Delta x_s = 0$ μm.

In the experiment, for every position of the microsphere, the maximum reflection as well as the corresponding azimuth angle $\theta_{max}$ are recorded, as shown in the black and blue curves in Fig. 4(b), respectively. The two vertical red dot-dash lines in Fig. 4(b) mark the two edges of the microsphere. A relatively high coupling efficiencies can be achieved in a large range of ~ 15 μm, in what scope the most energy of the incident Gaussian beam illuminates on the microsphere. It is worth noting that the reflection curve deviates from the mirror symmetry relative to the center of the microsphere ($\Delta x_s \sim 0$). This deviation may be attributed to the misalignment between the incident Gaussian beam and the $z$ axis. In addition, there is also a dip at $\Delta x_s \sim -3$ μm, which is because the coupling efficiency is recorded by the maximum intensities in the reflection spectrum of the doublet peak structure (see Fig. 1(b)). In most cases one peak in the doublet is much higher than the other one, then the maximum intensities well represents the coupling efficiencies of the splitting mode. However, the doublet peaks have nearly equal height at $\Delta x_s \sim -3$ μm, therefore get a much smaller apparent coupling efficiency. The azimuth angle $\theta_{max}$ shows an approximate linear relationship with $\Delta x_s$ (blue curve in Fig. 4(b)). A same deviation respect to the center is appear in the curve, which is also caused by the misalignment of the Gaussian beam.

## 3. CONCLUSION

To summarize, we have demonstrated a compact scheme to efficiently couple light to a high-Q WGM resonator to form a fiber based microprobe. The incident laser beam is focused by the microsphere itself as well as the GRIN lens, which is also used to collect the resonant reflection signal. With the help of the Purcell effect and the two-step focusing technique, the excitation efficiency of a high $Q$ factor WGM as high as 16.8% has been achieved. The proposed coupling method is robust to the mechanical vibration between the microsphere and the incident probe beam, and both the probe light and the reflection resonant signal are guided by the same single-mode fiber, making the whole coupling system an excellent fiber based WGM microprobe for a variety of applications, including lasing, sensing and imaging. By packaging the microsphere, the nanotip, the GRIN lens and the fiber together with low-refractive index glue [23] in the future, the WGM microprobe will open up a new direction for a variety of sensing/imaging applications, such as nanoparticle/biomolecule sensing, resonant acoustic imaging, and in-situ monitoring of dynamic chemical reactions.

# REFRENCES

1. L. Yang and K. J. Vahala, "Gain functionalization of silica microresonators," Opt. Lett. **28**, 592-594 (2003).
2. L. Yang, D. K. Armani, and K. J. Vahala, "Fiber-coupled erbium microlasers on a chip," Appl. Phys. Lett. 83, 825–826 (2003).
3. L. He, Ş. K. Özdemir, and L. Yang, "Whispering gallery microcavity lasers," Laser Photon. Rev. 7, 60–82 (2013).
4. X.-F. Jiang, Y.-F. Xiao, C.-L. Zou, L. He, C.-H. Dong, B.-B. Li, Y. Li, F.-W. Sun, L. Yang, and Q. Gong, "Highly Unidirectional Emission and Ultralow-Threshold Lasing from On-Chip Ultrahigh-Q Microcavities," Adv. Mater. 24, OP260-OP264 (2012).
5. X.-F. Jiang, C.-L. Zou, L. Wang, Q. Gong, and Y.-F. Xiao, "Whispering-gallery microcavities with unidirectional laser emission," Laser Photon. Rev. 10, 40-61 (2016).
6. M. Aspelmeyer, T. J. Kippenberg, and F. Marquardt, "Cavity optomechanics," Rev. Mod. Phys. 86, 1391–1452 (2014).
7. F. Monifi, J. Zhang, Ş. K. Özdemir, B. Peng, Y. Liu, F. Bo, F. Nori, and L. Yang, "Optomechanically induced stochastic resonance and chaos transfer between optical fields," Nat. Photonics 10, 399–405 (2016).
8. X. Jiang, M. Wang, M. C. Kuzyk, T. Oo, G.-L. Long, and H. Wang, "Chip-based silica microspheres for cavity optomechanics," Opt. Express 23, 27260–27265 (2015).
9. F. Vollmer, D. Braun, A. Libchaber, M. Khoshsima, I. Teraoka, and S. Arnold, "Protein detection by optical shift of a resonant microcavity," Appl. Phys. Lett. 80, 4057–4059 (2002).
10. J. Zhu, Ş. K. Özdemir, Y.-F. Xiao, L. Li, L. He, D.-R. Chen, and L. Yang, "On-chip single nanoparticle detection and sizing by mode splitting in an ultrahigh-Q microresonator," Nat. Photonics 4, 46–49 (2010).
11. L. Shao, X.-F. Jiang, X.-C. Yu, B.-B. Li, W. R. Clements, F. Vollmer, W. Wang, Y.-F. Xiao, and Q. Gong, "Detection of single nanoparticles and lentiviruses using microcavity resonance broadening," Adv. Mater. 25, 5616–5620 (2013).
12. V. R. Dantham, S. Holler, C. Barbre, D. Keng, V. Kolchenko, and S. Arnold, "Label-free detection of single protein using a nanoplasmonic-photonic hybrid microcavity, " Nano Lett. 13, 3347–3351 (2013)
13. Y. Zhi, X.-C. Yu, Q. Gong, L. Yang, and Y.-F. Xiao, " Single nanoparticle detection using optical microcavities," Adv. Mater. 29, 1604920 (2017).
14. S. H. Huang, S. Sheth, E. Jain, X.-F. Jiang, S. P. Zustiak, and L. Yang, "Whispering gallery mode resonator sensor for in situ measurements of hydrogel gelation," Opt. Express 26, 51-62 (2018)
15. B.-B. Li, Q.-Y. Wang, Y.-F. Xiao, X.-F. Jiang, Y. Li, L. Xiao, and Q. Gong, "On chip, high-sensitivity thermal sensor based on high-Q polydimethylsiloxane-coated microresonator," Appl. Phys. Lett. 96, 251109 (2010).
16. X. Xu, X. Jiang, G. Zhao, and L. Yang, "Phone-sized whispering-gallery microresonator sensing system," Opt. Express 24, 25905-25910 (2016).
17. L. Xu, X. Jiang, G. Zhao, D. Ma, H. Tao, Z. Liu, F. G. Omenetto, and L. Yang, "High-Q silk fibroin whispering gallery microresonator," Opt. Express 24, 20825-20830 (2016).
18. S. M. Spillane, T. J. Kippenberg, and K. J. Vahala, "Ultralow-threshold Raman laser using a spherical dielectric microcavity," Nature 415, 621–623 (2002).
19. T. J. Kippenberg, R. Holzwarth, and S. A. Diddams, "Microresonator-based optical frequency combs," Science 332, 555–559 (2011).
20. L. Yang, T. Carmon, B. Min, S. M. Spillane, and K. J. Vahala, "Erbium-doped and Raman microlasers on a silicon chip fabricated by the sol-gel process," Appl. Phys. Lett. 86, 1–3 (2005).
21. X.-F. Jiang, Y.-F. Xiao, Q.-F. Yang, L. Shao, W. R. Clements, and Q. Gong, "Free-space coupled, ultralow-threshold Raman lasing from a silica microcavity," Appl. Phys. Lett. 103, 101102 (2013).
22. Y. Yang, X. Jiang, S. Kasumie, G. Zhao, L. Xu, J. M. Ward, L. Yang, and S. N. Chormaic, "Four-wave mixing parametric oscillation and frequency comb generation at visible wavelengths in a silica microbubble resonator," Opt. Lett. 41, 5266 (2016).
23. G. Zhao, Ş. K. Ozdemir, T. Wang, L. Xu, E. King, G. L. Long, and L Yang, "Raman lasing and Fano lineshapes in a packaged fiber-coupled whispering-gallery-mode microresonator," Science Bulletin, 62, 875–878 (2017).
24. S. C. Hagness, D. Rafizadeh, S. T. Ho, and A .Taflove, "FDTD microcavity simulations: design and experimental realization of waveguide-coupled single-mode ring and whispering-gallery-mode disk resonators," J. Lightwave Technol. 15, 2154-2165 (1997).
25. M. Cai, O. Painter, and K. J. Vahala, "Observation of critical coupling in a fiber taper to silica-microsphere whispering gallery mode system," Phys. Rev. Lett. 85, 74-77 (2000).
26. J. C. Knight, G. Cheung, F. Jacques, and T. A. Birks, "Phase-matched excitation of whispering-gallery-mode resonances by a fiber taper," Opt. Lett. 22, 1129 (1997).
27. J. U. Nöckel and A. D. Stone, "Ray and wave chaos in asymmetric resonant optical cavities," Nature 1997, 385, 45-47 (1997).
28. J. Wiersig and M. Hentschel, "Combining Directional Light Output and Ultralow Loss in Deformed Microdisks," Phys. Rev. Lett. 100, 033901 (2008).
29. C. L. Zou, F. W. Sun, C. H. Dong, F. J. Shu, X. W. Wu, J. M. Cui, Y. Yang, Z. F. Han, and G. C. Guo, "High-Q and unidirectional emission whispering gallery modes: Principles and design," IEEE J. Sel. Top. Quantum 19, 9000406 (2013).
30. F. J. Shu, C. L. Zou, and F. W. Sun, "An Optimization Method of Asymmetric Resonant Cavities for Unidirectional Emission," J. Lightwave Technol. 31, 2994–2998 (2013).
31. C.-L. Zou, F.-J. Shu, F.-W. Sun, Z.-J. Gong, Z.-F. Han, and G.-C. Guo, "Theory of free space coupling to high-

Q whispering gallery modes," Opt. Express 21, 9982-9995 (2013).
32. L. Shao, L. Wang, W. Xiong, X.-F. Jiang, Q.-F. Yang, and Y.-F. Xiao, "Ultrahigh-Q, largely deformed microcavities coupled by a free-space laser beam," Appl. Phys. Lett. 103, 121102 (2013).
33. Z.-P. Liu, X.-F. Jiang, Y. Li, Y.-F. Xiao, L.Wang, J.-L. Ren, S.-J. Zhang, H. Yang, and Q. Gong, "High-Q asymmetric polymer microcavities directly fabricated by two-photon polymerization," Appl. Phys. Lett. 102, 221108 (2013).
34. Q.-F. Yang, X.-F. Jiang, Y.-L. Cui, L. Shao, and Y.-F. Xiao, "Dynamical tunneling-assisted coupling of high-Q deformed microcavities using a free-space beam," Phys. Rev. A 88, 023810 (2013).
35. F.-J. Shu, C.-L. Zou, and F.-W. Sun, "Dynamic process of free space excitation of asymmetric resonant microcavity," J. Lightwave Technol. 31, 1884-1889 (2013).
36. X. Jiang, L. Shao, S.-X. Zhang, X. Yi, J. Wiersig, L. Wang, Q. Gong, M. Lončar, L. Yang, and Y.-F. Xiao, "Chaos-assisted broadband momentum transformation in optical microresonators," Science 358, 344-347 (2017).
37. Y.-C. Liu, Y.-F. Xiao, X.-F. Jiang, B.-B. Li, Y. Li, and Q. Gong, "Cavity-QED treatment of scattering-induced free-space excitation and collection in high-$Q$ whispering-gallery microcavities," Phys. Rev. A 85, 013843 (2012).
38. Q. Song and H. Cao, "Highly directional output from long-lived resonances in optical microcavity," Opt. Lett. 36, 103–105 (2011).
39. J. Zhu, S. K. Ozdemir, H. Yilmaz, B. Peng, M. Dong, M. Tomes, T. Carmon, and L. Yang, "Interfacing whispering-gallery microresonators and free space light with cavity enhanced Rayleigh scattering," Sci. Rep. 4, 6396 (2014).